\documentclass[fleqn,usenatbib]{mnras}

\usepackage{tabularx}
\usepackage{comment}
\usepackage{url}
\usepackage{graphicx}
\usepackage[T1]{fontenc}
\usepackage{ae,aecompl}
\usepackage{booktabs}
\usepackage{gensymb}
\usepackage{float}
\usepackage{lmodern}
\usepackage{hyperref}

\usepackage{newtxtext,newtxmath}
\usepackage{amsmath}
\usepackage{natbib}
\usepackage{afterpage}

\usepackage{multicol}
\usepackage{pdflscape}
\usepackage{arydshln}

\def\ltsim{\mathrel{\hbox{\rlap{\hbox{\lower3pt\hbox{$\sim$}}}\hbox{\raise2pt\hbox{$<$}}}}}
\def\gtrsim{\mathrel{\hbox{\rlap{\hbox{\lower3pt\hbox{$\sim$}}}\hbox{\raise2pt\hbox{$>$}}}}}

\title[Hard-Soft X-ray Variability of Mrk 421 \& 1ES 1959+650]{Correlated Short-Timescale Hard-Soft X-ray Variability of the Blazars Mrk 421 and 1ES 1959+650 using \textit{AstroSat}}
\author[Das \& Chatterjee]{
Susmita Das,$^{1}$\thanks{E-mail: susmita.rs@presiuniv.ac.in}
Ritaban Chatterjee$^{1}$
\\
$^{1}$School of Astrophysics, Presidency University, 86/1 College Street, Kolkata, West Bengal, India 700073.
}

\date{Accepted XXX. Received YYY; in original form ZZZ}

\pubyear{2023}

\begin{document}
\label{firstpage}
\pagerange{\pageref{firstpage}--\pageref{lastpage}}
\maketitle

\begin{abstract}
We study simultaneous soft ($0.7-7$ keV) and hard ($7-20$ keV) X-ray light curves at a total of eight epochs during $2016-2019$ of two TeV blazars Mrk 421 and 1ES 1959+650 observed by the SXT and LAXPC instruments onboard \textit{AstroSat}. The light curves are $45-450$ ks long and may be sampled with time bins as short as $600-800$ sec with high signal to noise ratio. The blazars show a harder when brighter trend at all epochs. Discrete cross-correlation functions indicate that the hard and soft X-ray variability are strongly correlated. The time lag is consistent with zero in some epochs, and indicates hard or soft lag of a few hours in the rest. In the leptonic model of blazar emission, soft lag may be due to slower radiative cooling of lower energy electrons while hard lag may be caused by gradual acceleration of the high energy electrons emitting at the hard X-ray band. Assuming the above scenario and the value of the Doppler factor ($\delta$) to be $10-20$, the hard and soft lags may be used to estimate the magnetic field to be $\sim 0.1$ Gauss and the acceleration parameter to be $\sim 10^4$ in the emission region. Due to the availability of the high time resolution ($\sim$ minutes to hours) light curves from \textit{AstroSat}, the value of the illusive acceleration parameter could be estimated, which provides a stringent constraint on the theories of particle acceleration in blazar jets.
\end{abstract}


\section{Introduction} \label{sec:intro}
Jets are ubiquitous in the Universe. They are present in active galactic nuclei (AGN), X-ray binaries, supernova remnants and young stellar objects. Blazars are a class of AGN, which contains a bright relativistic jet, the axis of which is aligned to the observer's line of sight within an angle ${\sim 10^{\circ}}$ \citep{bla78, urr95}. Due to its orientation, jet emission is relativistically beamed in the observer's frame \citep[e.g.,][]{ghise1993} causing the spectral energy distribution (SED) of blazars to be dominated by the jet emission. Jets emit non-thermal radiation over a wide range of the electromagnetic spectrum often from radio to very high energy $\gamma$-rays. 

SED of blazars consist of two broad peaks \citep{abdo10a,giommi12}: one at the IR to UV/X-ray wave bands and another at the keV to GeV energy range sometimes extending to TeV. The lower energy peak is due to synchrotron radiation by the relativistic electrons moving in the magnetic field present in the jet \citep{bregman1981,urry1982,impey1988,marscher1998,ghis17}. According to the ``leptonic'' model of jet emission, the higher energy peak is due to the up-scattering of the lower energy ``seed'' photons by the same electrons through the inverse-Compton (IC) process. The source of seed photons may be synchrotron radiation in the jet itself termed synchrotron self-Compton \citep[SSC;][]{koni1981,maras1992,sikora1994,bloom1996,mast1997,bott07} or may be external to the jet, i.e., from the dusty torus or the broad line region which is called the external Compton (EC) process \citep{dermer1992,ghise1998,blejo2000,bott10}. Alternatively, in the ``hadronic scenario,'' protons in the jet may be accelerated to ultra-relativistic energies and contribute to the higher energy emission via synchrotron process, proton-initiated cascades and the interaction of secondary particles with photons \citep{mucke01, mucke03, bottcher13,botta16,acker16}, which are able to give rise to minutes-to-days timescale variability. Hadronic models have been used to satisfactorily model the higher energy (GeV-TeV) part of the SED of HSP blazars, such as Mrk 421 \citep[e.g.,][]{abdo11}. They find that in the hadronic scenario the size of the emission region is small (few gravitational radius of the central black hole), the magnetic field is large ($\sim$50 Gauss) and required proton energy is up to $10^{18}$ eV. We note that hadronic models are relatively less explored due to the above stringent requirements, complexity of the processes involved including several secondary particles, and leptonic models with less stringent constraints being able to satisfactorily fit the observed SEDs of blazars of different classes at different brightness states. However, if detection of neutrinos from the direction of flaring blazars, such as in the case of the blazar TXS 0506+056 \citep{ice18}, becomes more commonplace and significant then the usage of hadronic models to fit blazar SEDs will become much more prominent.

According to the position of the synchrotron peak ($\nu_{sync}$) in the SED, blazars are divided into three classes \citep{pad1995,abdo10}: LSP ($\nu_{sync}<10^{14}$ Hz), ISP ($10^{14}<\nu_{sync}<10^{15}$ Hz) and HSP ($\nu_{sync}>10^{15}$ Hz), i.e., low, intermediate and high synchrotron peaked, respectively.
 
The flux of blazars varies at minutes to years timescale. The nature of the variability is similar to red noise, i.e., longer timescale variability has larger amplitude \citep[e.g.,][]{lawrence1987,chatt12}. The variability at weeks to months timescale is believed to be caused by the crossing of shock waves through the emission region which energizes the quiescent particles \citep[e.g.,][]{marscher1985,marscher2016, bottcher19}. However, sometimes there is significant variability (factor of a few to even an order of magnitude) at minutes to days timescale, particularly at the X-ray and $\gamma$-ray energies. Sub-day timescale variability of blazars is caused by the acceleration and radiative cooling of the emitting particles \citep[e.g.,][]{zhang06,bottcher19}. While emission variability of blazars at multiple wave bands has often been probed to understand the physics of jets \citep[e.g.,][]{marscher2016} the sub-day timescale fluctuations are not commonly explored because light curves with the required time resolution are often not available. That is because either the blazar is not bright enough to obtain high time resolution ($\sim$ minutes$-$hours) monitoring with sufficient signal to noise ratio or the telescope time commitment for such continuous stare is too high.

In this work, we have analyzed $\sim$ minutes to $\sim$ days timescale X-ray variability of two blazars, namely, Mrk 421 and 1ES 1959+650, obtained by \textit{AstroSat} \citep{agrawal06,kpsing14} to probe the acceleration and cooling timescale of the emitting high energy particles. 

\subsection{Mrk 421}
Mrk 421 is one of the closest blazars at redshift $z=0.031$ \citep{piner1999}. It is an HSP blazar with the synchrotron peak of its SED at X-ray frequencies and the high energy peak at $\sim$ 50 GeV \citep[e.g.,][]{abdo11,ban19}. It was the first extragalactic source to be detected at TeV energies \citep{punch1992}. Being one of the brightest few blazars in the X-ray wave band, Mrk 421 has been studied by many authors using a number of X-ray telescopes as well as other multi-wavelength observations \citep[e.g.,][]{gai1996,kataoka01,tanihata04,cui04,fossati08,lichti08,horan09,kat10,tluc10,balokovic16,das21,marko22}. Several authors have found a strong correlation between the hard and soft X-ray variability with time lag consistent with zero \citep[e.g.,][]{sembay02,pandey17,zhongli19}. While the hard-soft time lag was consistent with zero in the 7-day-long \textit{ASCA} observation of Mrk 421 in 1998, significant hard/soft lag ($\sim$ hours) was found in shorter-timescale flares within those light curves \citep{takahashi1996,taka2000,tanihata01}. \citet{brinkmann03,brinkmann05} analyzed multiple \textit{XMM-Newton} light curves from different epochs to find strong correlation with non-zero time lag in some of those cases. Using long-term X-ray observations during 2000-2015 by \textit{Chandra} X-ray telescope, \citet{aggrawal18} found strong correlation with no significant time lag between the hard and soft X-ray variability. On the other hand, \citet{fossati2000,lichti08} found a hard lag (hard X-ray variability lagging those at the soft X-rays) in Mrk 421. Hard lag was also found with \textit{BeppoSAX} observations in 1998 \citep{zhang02} and the hard lag had a decreasing trend with a decrease of the energy difference between the hard-soft energy band pairs used for correlation.
 
\subsection{1ES 1959+650}
1ES 1959+650 is a relatively nearby ($z=0.048$) HSP blazar \citep{perlman1996}. It was one of the first few blazars to be detected at the TeV energies \citep{catanese1998,nishiyama1999,horns03}. 
The most recent TeV observation has been by the \textit{MAGIC} telescope \citep{magic20}. It was first discovered at the radio frequencies with the \textit{Green Bank Telescope} \citep{gregory1991,becker1991}. Later, it was observed at the optical wavelengths, in which it exhibited large-amplitude and fast variability \citep{villata2000}. The first detection at X-ray energies was by the \textit{Einstein-IPC} \citep{elvis1992}. Subsequently, it was observed by \textit{ROSAT} in 1996, by \textit{BeppoSAX} in 1997, and by \textit{Rossi X-Ray Timing Explorer (RXTE)} in 2000 \citep{beckmann02,giebels02}. More detailed studies of its X-ray and multi-wavelength spectra have been carried out by many authors \citep[e.g.,][]{tagli03,kraw04,perlman05,massaro08,bottacini10,kapa16,kapa216,kaur17,patel18,kapa18}. Recently, the X-ray spectra obtained from \textit{AstroSat}'s SXT and LAXPC instruments during a flaring event in 2017 have been analyzed and modeled in details \citep{zahir21,chandra21}. In 1ES 1959+650, hard and soft X-ray cross-correlation has not been explored too often. \citet{pandey17} did not find any significant intra-day variability in two \textit{NuSTAR} light curves of 1ES 1959+650 having a duration of $\sim$30 ks during 2014. Using those data, \citet{bhatta18} found strong correlation between the soft and hard X-rays with time lag consistent with zero. 

In both these HSP blazars, the synchrotron peak is at the $0.1-10$ keV range, which is observed by SXT and LAXPC. Therefore, in this work we study the hard and soft X-ray variability of these two blazars using \textit{AstroSat} data to explore the fluctuations due to the highest energy electrons. Our goal is to study the cross-correlation precisely using well sampled short-term data.

In section 2, we describe the reduction of the \textit{AstroSat} data and present the hard and soft X-ray light curves of Mrk 421 and 1ES 1959+650. Then we calculate the hardness ratio from the hard and soft X-ray light curves and plot its time evolution on the hardness-intensity plane at different epochs. We carry out discrete cross-correlation of the hard and soft X-ray light curves along with rigorous and precise estimates of the time lag and its uncertainties, and the significance of the cross-correlation in section 3. Finally, in section 4, we discuss the results and their implications regarding the emission mechanism and physical properties of the jet.   

\begin{figure*}
\centering
\includegraphics[height=20cm,width=\textwidth]{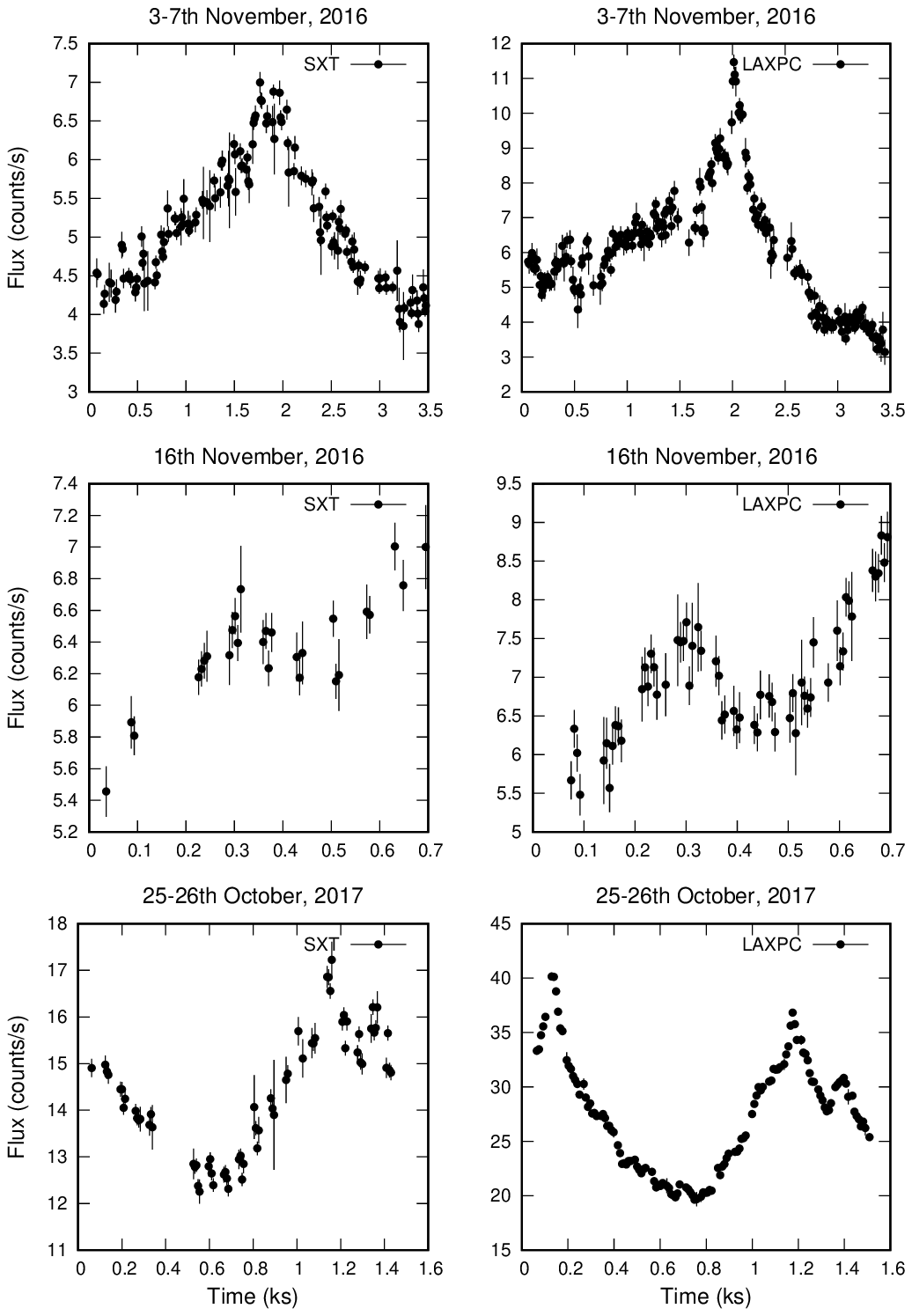} 
\caption{Black solid circles denote X-ray count rates of the blazar 1ES 1959+650 at multiple epochs during 2016-2017  observed by the Indian space telescope \textit{AstroSat}. The left and right panels show the X-ray light curves at $0.7-7$ keV from SXT and that at $7-20$ keV from LAXPC instrument, respectively, onboard \textit{AstroSat}.} \label{lc_1es1959}
\end{figure*}

\begin{figure*}
\centering
\includegraphics[height=22cm,width=\textwidth]{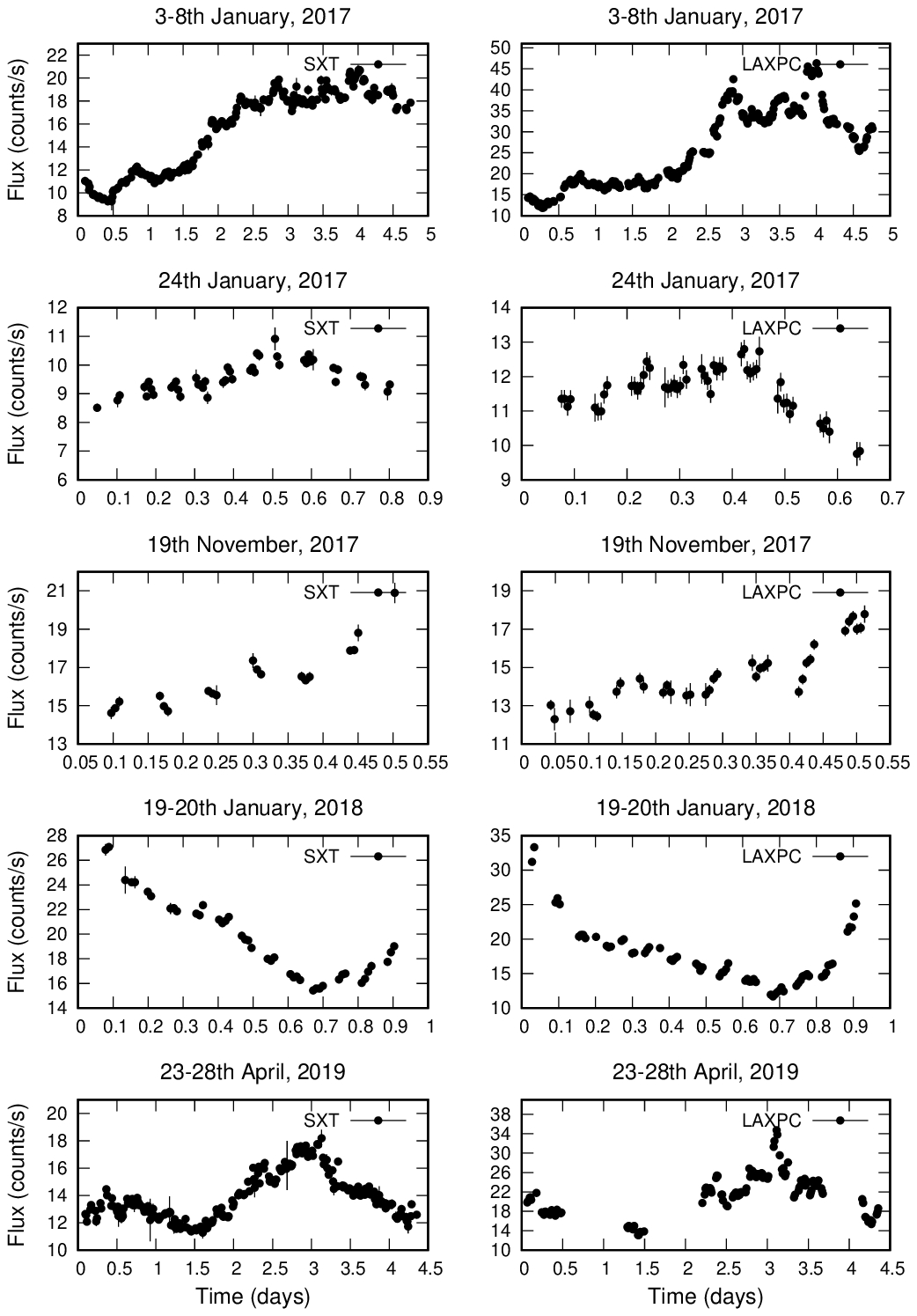} 
\caption{Black solid circles denote X-ray count rates of the blazar Mrk 421 at multiple epochs during 2016-2019 observed by the Indian space telescope \textit{AstroSat}. The left and right panels show the X-ray light curves at $0.7-7$ keV from SXT and that at $7-20$ keV from LAXPC instrument, respectively, onboard \textit{AstroSat}.} \label{lc_mrk421}
\end{figure*}

\section{\textit{AstroSat} Data}
\textit{AstroSat}, India's first space telescope dedicated to Astronomy, was launched on 2015 September 28 \citep{rao16,kpsing14}. It has five payloads, which are able to observe simultaneously at a range of wave bands from UV to very high energy X-rays (150 keV). All \textit{AstroSat} data are publicly available in its online data repository\footnote{\url{https://astrobrowse.issdc.gov.in/astro_archive/archive/Home.jsp}} after the 1-year proprietary period. 
 
 \begin{table*}
    \centering
    \caption{Parameters of the light curves presented in this work.}
    {\renewcommand{\arraystretch}{1.5}
    \begin{tabular}{|c|c|c|c|c|c|}
    \hline
    Source & Observation Date (UTC) & Start Time (MJD) & Stop Time (MJD) & SXT Exposure (ks) & LAXPC Exposure (ks) \\
    \hline
    1ES1959+650  &2016-11-03T20:47:00 &57695.9  &57699.1 &79.6 &73.2 \\
         &2016-11-16T10:29:46 &57708.4 &57709.1 &15.8 &31.7 \\
         &2017-10-25T02:28:50 &58051.1 &58052.6 &35.1 &85.7 \\
         \hdashline
    Mrk 421  &2017-01-03T13:31:35 &57756.5 &57761.2 &104.9 &168 \\
         &2017-01-24T07:34:24 &57777.3 &57778.1 &20.1 &31.2 \\
         &2017-11-19T07:38:51 &58076.3 &58076.9 &7.3 &18.3 \\
          &2018-01-19T23:16:46 &58138.0 &58139.0 &25.1 &36.9 \\
          &2019-04-23T20:25:13 &58596.9 &58601.3 &104.7 &166 \\
          \hline
         
    \end{tabular}}
    \label{tab:my_label}
\end{table*}

\subsection{Soft X-Ray Telescope (SXT)}
We have obtained the soft X-ray ($0.7-7$ keV) light curves from the Soft X-ray Telescope (SXT), which is an imaging telescope onboard \textit{AstroSat} \citep{kpsing16,kpsing17}.
The Level-1 data from SXT are stored in FITS format in the \textit{AstroSat} data archive. We generate Level-2 data from Level-1 by running \texttt{sxtpipeline} tool provided by SXT data analysis package (\textit{AS1SXTLevel2-1.3}). The Level-2 data consist of cleaned event files of each orbit. The South Atlantic Anomaly (SAA) and Earth occultation effects are avoided by using the data from the `good time intervals.' The cleaned event files of individual orbits are then merged together by \texttt{sxtpyjuliamerger}. We use the \texttt{xselect} tool of \textit{HEASoft} to extract the image, light curve and spectrum from the merged cleaned event file. We select the source region as a circle of radius $16'$ centered at the point source. To obtain the background light curve, we use SXT background spectra SkyBkg{\_}comb{\_}EL3p5{\_}Cl{\_}Rd16p0{\_}v01.pha to get constant background counts in the desired energy range.
The final light curves are obtained using the tool \texttt{lcurve}, which is also used to bin the data. 

\subsection{Large Area X-ray Proportional Counter (LAXPC)}
 We have obtained the hard X-ray ($7-20$ keV) light curves from the Large Area X-ray Proportional Counter (LAXPC) onboard \textit{AstroSat}. In LAXPC, three co-aligned proportional counters detect X-ray photons between $3-80~ keV$ with a large effective area of $\sim 6000$ cm$^{2}$ \citep{yadav16,antia17}. The three detectors, LAXPC10, LAXPC20 and LAXPC30, have a field of view $\sim1^{\circ} \times 1^{\circ}$. They record the signal in Event analysis (EA) and Broad Band Counting (BB) mode which are stored as Level-1 data. Currently, only LAXPC20 is working nominally \citep{antia21}. Therefore, we only use the data from LAXPC20 and make light curve and spectrum of the source and the background using the \textit{LAXPCsoftware:~format(A)} package available in the \textit{AstroSat Science Support Cell (ASSC)} website\footnote{\url{http://astrosat-ssc.iucaa.in}/}. To extract the background effects we use the \textit{``Faint source background estimation"} scheme \citep{ranjeev21} because blazars are considered to be faint source for \textit{AstroSat} unless they are in a particularly bright state. This scheme is applicable for sources, in which the count rate at $50-80$ keV is less than $0.25$ counts\,s$^{-1}$. We use the \texttt{lcmath} and \texttt{lcurve} tools to obtain the final background subtracted light curves with different lengths of time bins.

\subsection{Light Curves}
We show the soft ($0.7-7$ keV from SXT) and hard ($7-20$ keV from LAXPC) X-ray light curves of the blazars Mrk 421 and 1ES 1959+650 at a total of eight epochs during $2016-2019$ obtained by the above process in Figures \ref{lc_1es1959} and \ref{lc_mrk421}. 
Our primary goal is to study the precise cross-correlation of the hard-soft X-ray variability. Therefore, we limit the lower end of the hard X-ray energy range to  $7$ keV for the LAXPC light curve to avoid any overlap with that of the SXT light curve in order to keep the soft and hard energy bands completely separate. We set the upper limit of the same to $20$ keV because the LAXPC background noise starts to dominate above that energy \citep[e.g.,][]{hota21}. Furthermore, we use a lower limit of 0.7 keV for our SXT light curves to avoid certain instrumental features \citep[e.g.,][]{zahir21} that are present at energies lower than that. The length of the light curves are $45-450$ ks with time bins as short as $600$ s to $800$ s. We use the shortest bins for which the average relative uncertainty is less than $10\%$.


\subsection{Hardness Intensity Diagrams}
\begin{figure*}
\centering
\includegraphics[height=19cm,width=\textwidth]{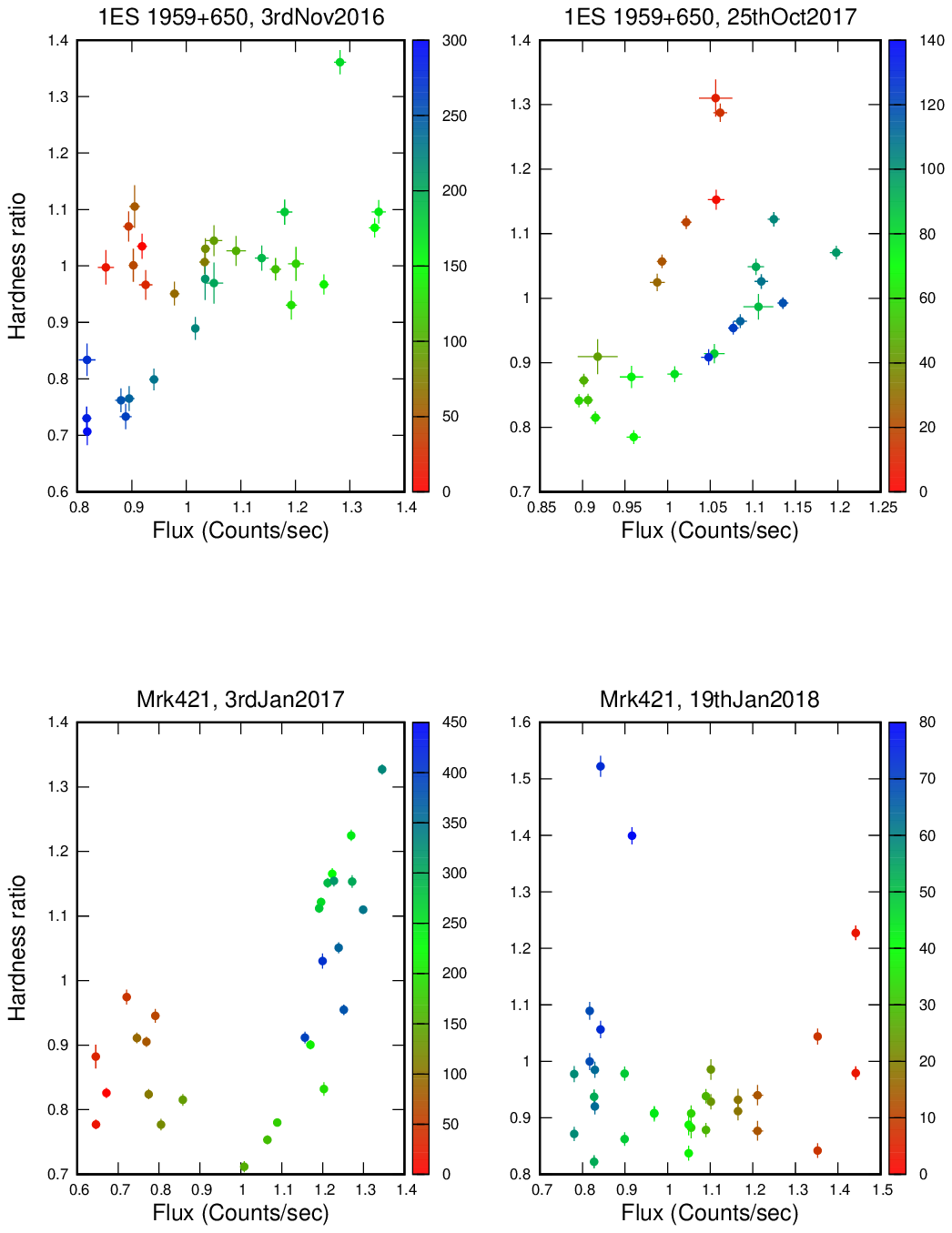} 
\caption{Hardness ratio ($7-20$ keV/ $0.7-7$ keV) vs photon flux ($0.7-7$ keV) of the two blazars 1ES 1959+650 and Mrk 421 at different epochs during 2016-2019. The color bars denote the time evolution in ks. } \label{hid_all_photon}
\end{figure*}

Hardness ratio is quantified here by the ratio of the count rate at the hard X-ray energy band ($7-20$ keV) to that at the soft X-ray energy band ($0.7-7$ keV). Hardness ratio is related to the spectral nature of the source. Figure \ref{hid_all_photon} shows the hardness-intensity diagram (HID), i.e., hardness ratio versus the observed photon flux at $0.7-7$ keV of Mrk 421 and 1ES 1959+650 for two long exposures of each. We rebin the light curves to make any existing pattern in the figure clearly visible, and normalize the hardness and intensity values by their mean to facilitate comparison between epochs. We can see that the hardness ratio increases with increasing flux in all cases, which implies a harder when brighter trend. In some epochs, the evolution in the HID exhibits a loop pattern, e.g., a clockwise and counter clockwise loop structure present in the 2018 epoch of Mrk 421 and 2017 epoch of 1ES 1959+650, respectively while in the HID plot of 1ES 1959+650 in the 2016 epoch no clear pattern is present. 

\begin{figure*}
\centering
\includegraphics[height=20cm,width=\textwidth]{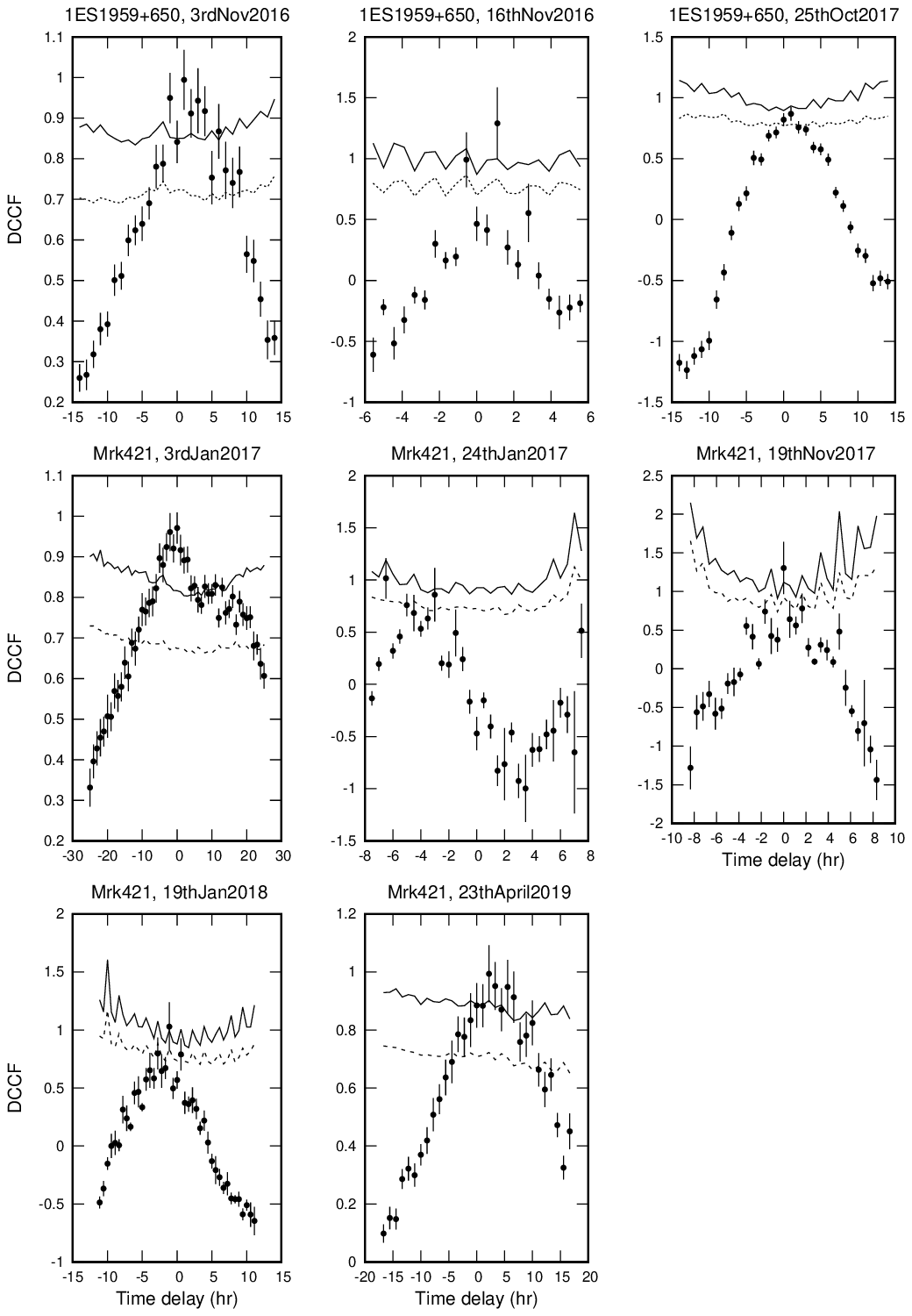}
\caption{Black solid circles denote the discrete cross-correlation functions of the soft ($0.7-7$ keV) and hard ($7-20$ keV) X-ray variability of the blazars 1ES 1959+650 and Mrk 421 during the epochs shown in Figures \ref{lc_1es1959} and \ref{lc_mrk421}. The dashed and solid lines indicate 95\% and 99\% significance levels, respectively. Positive values of time delay denote a hard lag, i.e., the hard band variability lags those at the soft band.}
\label{dcf_all}
\end{figure*}

\section{Cross-Correlation Analysis}
We cross-correlate the hard and soft X-ray light curves of both blazars for each of the epochs using the discrete cross-correlation function \citep[DCCF;][]{edelson1988} method. The cross-correlation function between the hard and soft X-ray variability may indicate whether the fluctuations at those two bands are causally related. Absence or presence of non-zero time lag between the variability may provide information about the geometric and physical parameters of the emission region, e.g., magnetic field and energy of the emitting electrons.

\subsection{Uncertainty of Time Lag}
A unique aspect of the light curves we are using here is that they provide X-ray variability information at a very high time resolution, e.g., sub-hour in some cases. Therefore, if there is a sub-hr time delay between the hard and soft X-ray variability that may be detected with high significance from the cross-correlation analysis of those light curves. However, the uncertainty of such time delay need to be computed carefully in order to ensure an accurate interpretation of the results. For that purpose, we use the model independent Monte-Carlo method, namely, flux randomization and random subset selection or FR-RSS \citep{peterson1998,sun18} to determine the time delay and its uncertainties.

We modulate the flux values by adding noise to each data point drawn from a normal distribution with a mean equal to the mean observed uncertainty. We select a random subset of each of the light curves consisting of $\sim$ 63.3\% of the data points. We make many realizations in both bands using the above flux randomization and random subset selection, and compute the cross-correlation function of each such pair. It generates a distribution of cross correlation results. We find the peak value of the correlation function and the corresponding time lag, termed, ``peak lag'' from the median of the distribution. Furthermore, we generate a distribution of ``centroid lags'' by calculating the mean time lag weighted by the corresponding correlation coefficients of the points, which are above $80\%$ of the peak value. We find the centroid lag from the median of that distribution. 

\subsection{Significance of Correlation}
The cross-correlation results of a pair of red noise light curves may be artificially high or low due to gaps in the data, irregular sampling or just by random chance associated with the stochastic nature of the variability. In order to test the significance of the cross-correlation results we simulate 100 each of hard and soft X-ray light curves from their power-spectral density (PSD) using the algorithm by \citet{timmer1995}. We use the soft and hard X-ray power-spectral shape and slope from \citet{chatt18} for both sources.

We cross-correlate the light curves simulated in the above method and find the 99 and 95 percentile levels of the distribution of cross-correlation values for each step of the time delays. We determine the significance of the DCCF values we obtain for the observed light curves by its comparison to the 99 and 95 percentile values obtained above. This provides an estimate of the probability that the value of the DCCF we obtain is high by chance and in turn provides a quantitative measure of its significance. 

\begin{table}
\centering
\caption{Time Lag and its uncertainty of the correlated soft ($0.7-7$ keV) and hard ($7-20$ keV) X-ray variability of the blazar 1ES 1959+650 at three different epochs during $2016-17$. Positive values of time delay denote a hard lag, i.e., the hard band variability lags those at the soft band. The uncertainties have been computed using the FR-RSS technique (see text).}\label{lag1959}
{\renewcommand{\arraystretch}{1.6}
\begin{tabular}{|c|c|c|}
\hline
Observation Date & Peak lag (hr) & Centroid lag (hr)  \\
\hline
2016 November 3-7 & $1.63^{+0.3}_{-0.21}$ & $0.99^{+0.48}_{-0.41}$	\\
2016 November 16 & $-0.02^{+0.66}_{-0.42}$ & $0.09^{+0.53}_{-0.45}$ \\
2017 October 25-26 & $1.17^{+0.8}_{-0.18}$ & $1.98^{+0.36}_{-0.33}$	\\
\hline
\end{tabular}}\quad
\label{lags_1959}
\end{table}

\begin{table}
\centering
\caption{Time Lag and its uncertainty of the correlated soft ($0.7-7$ keV) and hard ($7-20$ keV) X-ray variability of the blazar Mrk 421 at five different epochs during $2016-19$. Positive values of time delay denote a hard lag, i.e., the hard band variability lags those at the soft band. The uncertainties have been computed using the FR-RSS technique (see text).}\label{lag421}
{\renewcommand{\arraystretch}{1.6}
\begin{tabular}{|c|c|c|}
\hline
Observation Date & Peak lag (hr) & Centroid lag (hr) \\
\hline			
2017 January 3-8 & $-0.05^{+0.45}_{-0.15}$ & $7.35^{+0.95}_{-0.76}$ \\
2017 January 24 & $-4.39^{+0.6}_{-0.93}$ & $-4.46^{+0.58}_{-0.52}$ \\
2017 November 19 & $0.71^{+0.57}_{-0.51}$ & $0.65^{+0.30}_{-0.28}$ \\
2018 January 19-20 & $-0.9^{+0.48}_{-0.54}$ & $-2.07^{+2.20}_{-0.38}$\\
2019 April 23-28 & $2.14^{+1.44}_{-1.53}$ & $2.86^{+0.58}_{-0.44}$ \\
\hline
\end{tabular}}
\label{lags_mrk421}
\end{table}

\subsection{Results}
The discrete cross-correlation function of each pair of hard and soft X-ray light curves of Mrk 421 and 1ES 1959+650 for all of the epochs are shown in Figure \ref{dcf_all}. The solid and dotted lines denote the 99\% and 95\% confidence levels calculated in the method discussed in the previous section. We find that the hard and soft X-ray variability are strongly correlated with at least 95\% confidence level in all cases.

If the cross-correlation function is symmetric in shape around its maximum then the peak and centroid lag values are approximately the same. However, in some cases the CCFs are not symmetric leading to a significant difference between the peak and centroid lag. Therefore, we have reported both in Tables \ref{lags_1959} and \ref{lags_mrk421}. 

In a few cases, e.g., in the epochs 2016 November 16 of 1ES 1959+650 and 2018 January 19-20 of Mrk 421, the time lags (centroid) are consistent with zero within their uncertainties. Some other light curves exhibit a significant lag of $\sim$hr, i.e., variability at the hard band lagging those at the soft X-rays or \textit{vice versa}, termed as hard lag and soft lag, respectively. In the convention we use, a positive value of the time delay indicates hard lag and a negative time delay denotes a soft lag. For example, 1ES 1959+650 in the 2017 epoch and Mrk 421 in the 2017 January 3-8 epoch exhibit significant hard lag (centroid) while Mrk 421 shows a significant soft lag (centroid) during the 2017 January 24 epoch, beyond the corresponding uncertainty values. 
    
\section{Discussion \& Conclusion}
Here we have presented the hard and soft X-ray light curves of two HSP blazars, namely, Mrk 421 and 1ES 1959+650 at different epochs during $2016-2019$ observed by the SXT and LAXPC instruments onboard \textit{AstroSat}.

In the hardness-intensity diagrams the sources exhibit a harder when brighter trend. It has been previously reported in several observations of Mrk 421 and 1ES 1959+650 \citep[e.g.][]{sembay02,brinkmann03,ravasio04,chandra21}. The evolution in the hardness-intensity plane follows a clockwise or anti-clockwise path in some epochs. Such pattern indicates, for example, the higher energy variations propagate to lower energy or \textit{vice versa} causing soft or hard lag, respectively \citep{zhang02,zhang06}. 

In the leptonic model of emission from the blazar jets, the X-rays from the HSP blazars are produced by synchrotron radiation from the relativistic electrons in the jet \citep[e.g.,][]{abdo10,kapa18,giommi21}. Here we use the SXT and LAXPC energy ranges, i.e., $0.7-7$ keV and $7-20$ keV, respectively which include the lower energy peak of the SED as well as its decaying part beyond the peak energy. We have shown the spectra of both blazars at all the epochs discussed here along with the best-fit broken power-law model and spectral parameters in Appendix A. It shows that the $0.7-20$ keV spectra at all the epochs are consistent with a single component supposedly due to synchrotron radiation and no significant additional component is present in any of the epochs. A more detailed analyses of the multi-epoch spectra will be carried out in a future paper.
Several other authors, using various samples and observations, have concluded that in the HSP blazars, including the two discussed here, the decaying part of the synchrotron peak is located at the hard X-ray band (e.g., $\sim10-20$ keV) and the IC component does not contribute below 50 keV or even higher \citep{lichti08,abdo11,bhatta18,chandra21,zahir21,marko22,mondal22}. Therefore, the X-ray emission analyzed here is considered to be due to synchrotron emission only, and is, in fact, located at or near the lower energy peak of the SED.
Hence, the electrons responsible for generating the X-rays are at the peak of their energy distribution. The cooling timescales of such high energy electrons are short ($\sim$hr). Therefore, it is assumed that they are physically located in a compact region and undergo acceleration and cooling at approximately the same time. That is consistent with the strong correlation with zero or very short time delay between the hard and soft X-ray variability as we have found here.

There may be short but non-zero time delay caused by cooling and acceleration processes \citep[e.g.,][]{dermer09, chen11, lewis16}. For example, higher energy electrons cool faster in the synchrotron process because the synchrotron cooling timescale is inversely proportional to the energy of the emitting particles. Therefore, electrons producing hard X-rays will cool faster than those producing the soft X-rays causing the latter lagging the former, namely, a soft lag.

On the other hand, if the acceleration timescale is comparable or longer than that for cooling it may take a significant amount of time to energize the electrons enough so that they can emit synchrotron radiation at the hard X-ray energies, which will cause the hard X-ray variability to lag those at the soft X-rays, namely, a hard lag.

It is often assumed that in blazar jets the particles are accelerated by moving shocks through, e.g., a first order Fermi acceleration mechanism \citep{bland1987,kirk1998,kusunose2000,zhang2002a}. The acceleration timescale ($t_{acc}$) in that case is proportional to the final energy achieved by the particles while synchrotron cooling timescale ($t_{cool}$) is inversely proportional to the energy of the particle. The particles will reach the highest energy when $t_{acc} \simeq t_{cool}$, which indicates that below the highest energy, $t_{acc} << t_{cool}$ \citep{bottcher19}. For HSP blazars, the electrons that emit X-rays are at the peak of their energy distribution and we may assume $t_{acc} \simeq t_{cool}$. That is consistent with the observed result that we see zero time lag in some cases between the hard and soft X-ray variability while in some cases we do find a small amount of soft or hard lag. In particular, hard lag indicates the acceleration timescale is longer than our time resolution and we are able to probe it through our light curves and inter-band time delays.

The soft/hard lag may be expressed as below:
\begin{equation}
\tau_{soft}= t_{cool}(E_{soft}) - t_{cool}(E_{hard}) \label{softlag}
\end{equation}
\begin{equation}
\tau_{hard}= t_{acc}(E_{hard}) - t_{acc}(E_{soft}) \label{hardlag}
\end{equation}

The energy dependent acceleration and cooling timescales may be given as the following \citep{zhang2002a,ravasio04}:
\begin{equation}
t_{acc} (E) = 9.65 \times 10^{-2} (1+z)^{3/2} \xi B^{-3/2} \delta^{-3/2} E^{1/2} ~~\text{s} \label{acc_time}
\end{equation}
\begin{equation}
t_{cool} (E) = 3.04 \times 10^3 (1+z)^{1/2} B^{-3/2} \delta^{-1/2} E^{-1/2} ~~\text{s} \label{cool_time}
\end{equation} 
\\
Here, $E$, $z$, $B$ and $\delta$ are the energy of the emitted photon, redshift of the blazar, magnetic field at the emission region and the Doppler factor, respectively. $\xi$ is the so-called `acceleration parameter' or `acceleration efficiency'. 

Substituting Equations (\ref{acc_time}) and (\ref{cool_time}) into Equations (\ref{softlag}) and (\ref{hardlag}), one gets \citep[as in][]{kaza1998,zhang06}:
\begin{equation}
    B \delta^{1/3} = 209.91 \times\left(\frac{1+z}{E_s}\right)^{1/3}{\left[\frac{1-(E_s/E_h)^{1/2}}{\tau_{soft}}\right]}^{2/3} ~~\text{G}
\end{equation}
\begin{equation}
    B \delta \xi^{-2/3} = 0.21\times (1+z){E_h}^{1/3}{\left[\frac{1-(E_s/E_h)^{1/2}}{\tau_{hard}}\right]}^{2/3} ~~\text{G}
\end{equation}
\\
Here, $E_s$ and $E_h$ are the logarithmic mean energy of the corresponding soft and hard energy bands, respectively. Assuming a range of $\delta \sim 10-20$ we calculate the magnetic field in the emission region. In the case of Mrk 421, considering the soft lag (centroid lag) and the hard lag, we obtain the value of $B$ to be in the range $0.05-0.12$ G. Consequently, the acceleration parameter is $\xi \sim 10^4$. This value is consistent with that obtained by \citet{zhang02} using the observations of Mrk 421 in 1998. In the case of 1ES 1959+650 there is a non-zero time lag in 2017 which is a hard lag. From this time lag, we obtain $B \delta \xi^{-2/3} = 1.0 \times 10^{-3}$. If $B \sim 0.1$ G and $\delta = 10$, the value of $\xi$ is $\sim 10^4$. From theoretical modeling of the time-resolved X-ray spectra during a flare in 2017, \citet{zahir21} estimated the acceleration timescale in this source to be $2 \times 10^5$ s. The acceleration parameter $\xi$ is the ratio of the mean free path to the gyroradius $(\lambda/r_g)$ in the acceleration region 
\citep{inoue1996,baring1997,baring17}.
We obtain $\xi\gg1$ although in other astrophysical plasma, such as, the interstellar medium or supernova remnants $\xi \simeq 1$ \citep[e.g.,][]{inoue1996}. A large value of the acceleration parameter implies that the mean free path of scattering is large compared to the gyroradius ($r_g$). For relativistic electrons $r_g = \frac {\gamma m c^2}{e B}$. Assuming $B=0.1$ G, $r_g \sim 10^4 \gamma$ cm. That implies the mean free path of the electrons with $\gamma \sim 10^2$ will be $10^{10}$ cm for $\xi = 10^4$. If we assume the electrons are moving at a speed close to that of light, they will traverse $\sim 10^5 \lambda$ in an interval equal to the acceleration timescale $\sim 10^5$ s. Therefore, they will undergo $\sim 10^5$ scatterings in the acceleration region to be accelerated to the maximum energy ($\gamma \sim 10^5$) starting from $\gamma \sim 10^2$. That indicates the energy amplification in each scattering, assuming it is uniform at all values of initial energy, is $7 \times 10^{-3}$\%. If we make the same assumptions as above the electrons will undergo $10^3$ scatterings in order to reach $\gamma = 10^5$ from $\gamma = 10^4$. In that case, the energy amplification is $0.23$\% in each scattering. 
Due to the high time resolution light curves used in this work, we can put a meaningful constraint on the mechanism and efficiency of the acceleration of particles in the blazar jets.

\section{Data Availability}
The data utilized in this work are available in ISRO's online data repository.\\
Weblink:\url{https://astrobrowse.issdc.gov.in/astro_archive/archive/Home.jsp}\\
The software for AstroSat data reduction are available in the website of the \textit{AstroSat Science Support Cell (ASSC)}. \\
Weblink:\url{http://astrosat-ssc.iucaa.in}

\section{Acknowledgements} 
We thank the anonymous referee for suggestions that improved the manuscript. This work has used data from the Indian Space Science Data Centre (ISSDC) under the \textit{AstroSat} mission of the Indian Space Research Organisation (ISRO). We acknowledge the POC teams of the SXT and LAXPC instruments for archiving data and providing the necessary software tools. SD and RC thank ISRO for support under the \textit{AstroSat} archival data utilization program, and IUCAA for their hospitality and usage of their facilities during their stay at different times as part of the university associateship program. RC thanks Presidency University for support under the Faculty Research and Professional Development (FRPDF) Grant, RC acknowledges financial support from BRNS through a project grant (sanction no: 57/14/10/2019-BRNS) and thanks the project coordinator Pratik Majumdar for support regarding the BRNS project. We thank Jayashree Roy, Sunil Chandra, Ritesh Ghosh, Savithri Ezhikode, Gulab Dewangan and Ranjeev Misra for useful discussion regarding AstroSat data analysis.

\bibliographystyle{mnras}
\bibliography{correlation_mnras}

\appendix

\section{Multi-Epoch Spectra}
\begin{figure*}
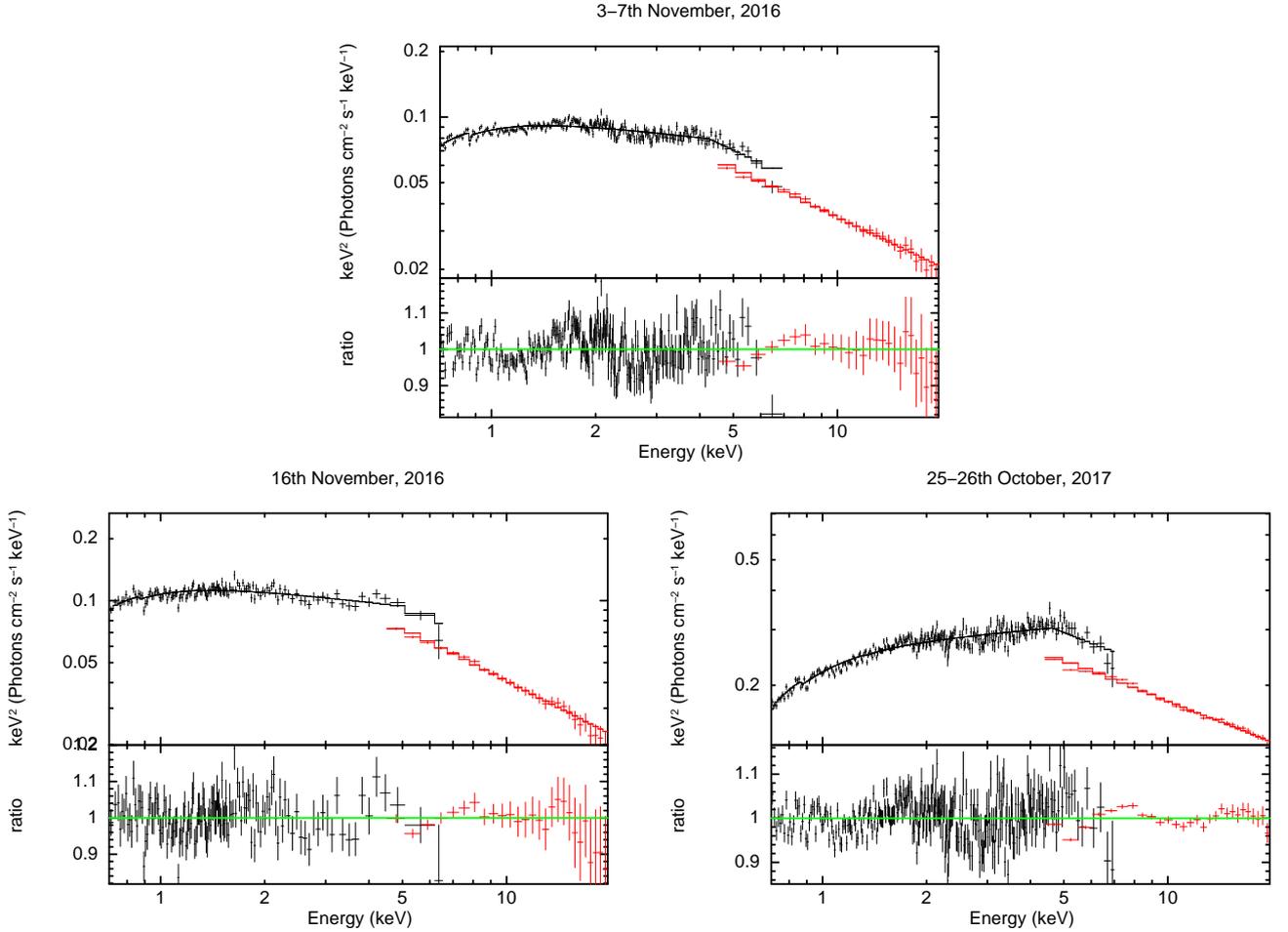

   \begin{minipage}{0.5\textwidth}
     \centering
     \includegraphics[width=.7\textwidth,angle=-90]{spec1959_0774_bknpow_0.7-20.eps}
     \end{minipage}
    \begin{minipage}{0.5\textwidth}
     \centering
     \includegraphics[width=.7\textwidth,angle=-90]{spec1959_0800_bknpow_0.7-20.eps}
  \end{minipage}\hfill
 \begin{minipage}{0.5\textwidth}
     \centering
     \includegraphics[width=.7\textwidth,angle=-90]{spec1959_1638_bknpow_0.7-20.eps}
\end{minipage}
   \caption{The black and red data points with error bars represent the SXT and LAXPC spectra, respectively, of 1ES 1959+650 during three epochs in 2016-17 while the best-fit broken power-law model is denoted by the black and red lines.}\label{spec_1959}
\end{figure*}

\begin{figure*}
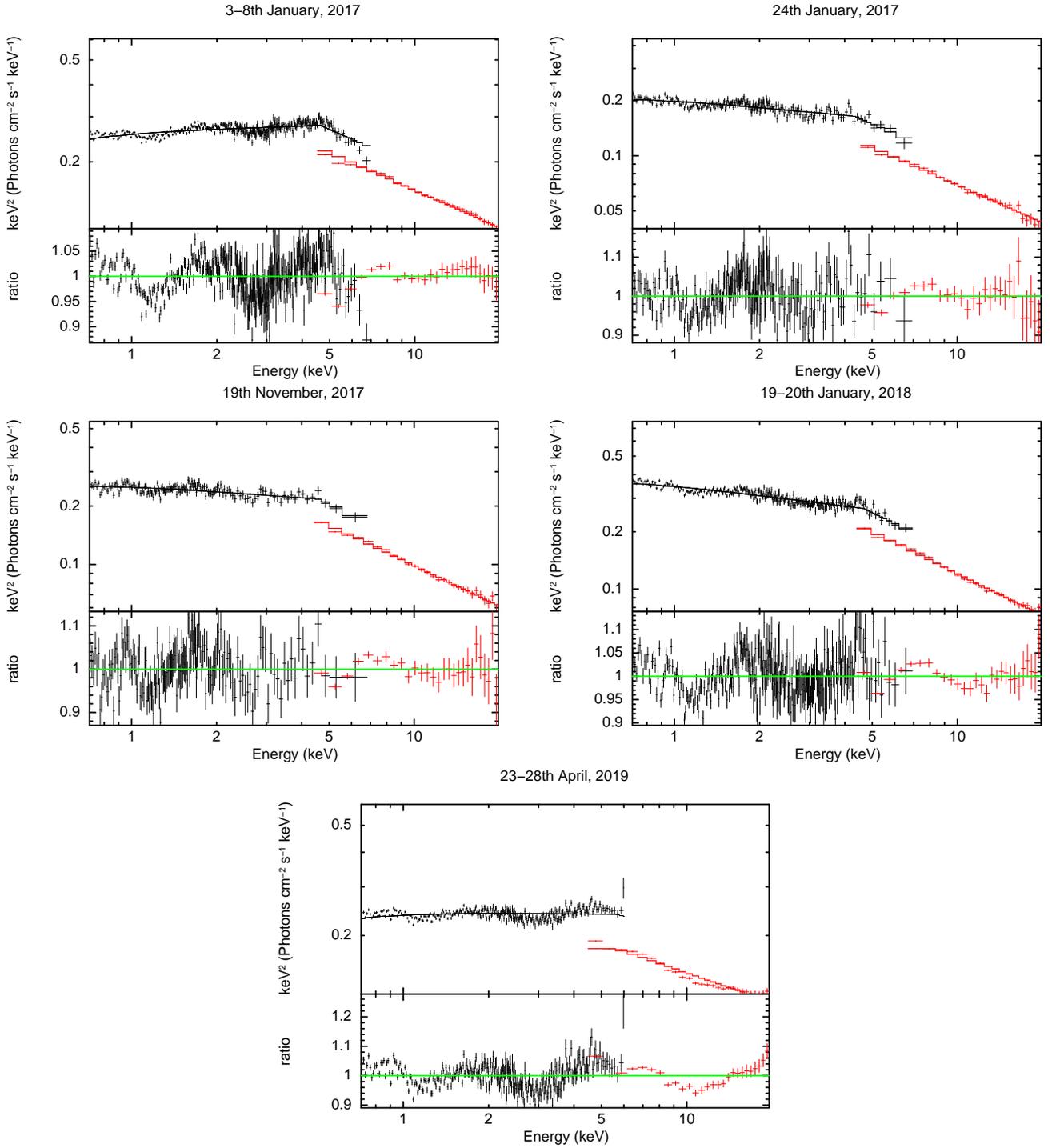

   \begin{minipage}{0.5\textwidth}
     \centering
     \includegraphics[width=.7\linewidth,angle=-90]{spec421_0948_bknpow_0.7-20.eps}
   \end{minipage}\hfill
      \begin{minipage}{0.5\textwidth}
     \centering
     \includegraphics[width=.7\linewidth,angle=-90]{spec421_0978_bknpow_0.7-20.eps}
   \end{minipage}\hfill
      \begin{minipage}{0.5\textwidth}
     \centering
     \includegraphics[width=.7\linewidth,angle=-90]{spec421_1704_bknpow_0.7-20.eps}
   \end{minipage}\hfill
      \begin{minipage}{0.5\textwidth}
     \centering
     \includegraphics[width=.7\linewidth,angle=-90]{spec421_1852_bknpow_0.7-20.eps}
   \end{minipage}\hfill
   \begin{minipage}{0.5\textwidth}
     \centering
     \includegraphics[width=.7\linewidth,angle=-90]{spec421_2856_bknpow_0.7-20.eps}
   \end{minipage}
   \caption{The black and red data points with error bars represent the SXT and LAXPC spectra, respectively, of Mrk 421 during five epochs in 2017-19 while the best-fit broken power-law model is denoted by the black and red lines.}\label{spec_421}
\end{figure*}

We carry out the spectral modeling of 1ES 1959+650 and Mrk 421 at all the epochs discussed in this work using \texttt{XSPEC} (version 12.10), a tool of \textit{HEASoft}. The \texttt{GRPPHA} command helps to group the source spectrum with response and background files. We group all channels by minimum $50$ counts per channel. For SXT we use background from blank-sky observations, i.e., {SkyBkg{\_}comb{\_}EL3p5{\_}Cl{\_}Rd16p0{\_}v01.pha} and the response {sxt{\_}pc{\_}mat{\_}g0to12.rmf} which are provided by SXT POC. We correct the given ARF {sxt{\_}pc{\_}excl00{\_}v04{\_}20190608.arf} with corresponding selected source region of radius $16'$ by using SXTARFModule. In LAXPC we group with the response of LAXPC20 and background which is extracted using the faint source background scheme. We fit SXT and LAXPC spectrum jointly in the energy range $0.7-20$ keV using the broken power-law model with Galactic absorption correction by \texttt{TBabs}. A constant factor representing the cross-normalization between the two instruments is fixed to 1 for SXT and kept as free parameter for LAXPC. We consider $3\%$ systematic error for combined fitting \citep[e.g.,][]{chandra21,hota21} as suggested by the instrument team. 
The best-fit parameters are listed in Table \ref{spec_fit} and the spectra with best-fit models are shown in Figures \ref{spec_1959}and \ref{spec_421}.

\begin{table*}
\centering
\caption{Best-fit parameters of the broken power-law model}\label{spec_fit}
\begin{tabular}{|l|l|l|l|l|l|l|l|} 
\hline
Source & Observation date & Cross-Normalization &$\Gamma_1$ & $\Gamma_2$ & $E_{\rm break}$ (keV) & $log_{10}$(Flux in erg\,cm$^{-2}$ \,s$^{-1}$) &$\chi^2/ d.o.f$  \\
 & &  Constant & & & & ~~ ($0.7-20$ keV) & \\
\hline
1ES 1959+650 & 2016, 3-7th Nov & 0.82 $\pm$ 0.02 & 2.19 $\pm$ 0.01 & 2.75 $\pm$ 0.05 & 4.25 $\pm$ 0.2 & -9.42 $\pm$ 0.03 & 1.05 \\
 & 2016, 16th Nov & 0.77 $\pm$ 0.04 & 2.19 $\pm$ 0.02 & 2.85 $\pm$ 0.05 & 5.09 $\pm$ 0.05 & -9.41 $\pm$ 0.6 & 0.88 \\
 &  2017, 25-26th Oct & 0.81 $\pm$ 0.02 & 1.91 $\pm$ 0.01 & 2.43 $\pm$ 0.02 & 4.63 $\pm$ 0.2 & -8.89 $\pm$ 0.1 & 1.1 \\ 
 \hline
 Mrk 421 & 2017, 3-8th Jan & 0.81 $\pm$ 0.02 & 1.96 $\pm$ 0.01 & 2.48 $\pm$ 0.02 & 4.63 $\pm$ 0.1 & -8.94 $\pm$ 0.08 & 1.1\\
 & 2017, 24th Jan & 0.74 $\pm$ 0.03 & 2.14 $\pm$ 0.01 & 2.67 $\pm$ 0.03 & 4.33 $\pm$ 0.3 & -9.13 $\pm$ 0.03 & 0.96 \\
 & 2017, 19th Nov & 0.77 $\pm$ 0.03 & 2.10 $\pm$ 0.01 & 2.68 $\pm$ 0.03 & 4.60 $\pm$ 0.35 & -9.00 $\pm$ 0.003 & 0.84 \\
 & 2018, 19-20th Jan & 0.79 $\pm$ 0.02 & 2.18 $\pm$ 0.01 & 2.73 $\pm$ 0.02 & 4.68 $\pm$ 0.18 & -8.89 $\pm$ 0.02 & 1.0 \\
 & 2019, 23-28th April & 0.75 $\pm$ 0.02 & 2.0 $\pm$ 0.006 & 2.35 $\pm$ 0.02 & 5.68 $\pm$ 0.38 & -8.97 $\pm$ 0.04 & 1.25 \\
\hline
\end{tabular}
\end{table*}

\label{lastpage}
\end{document}